\documentclass{webofc}
\usepackage[varg]{txfonts}   % Web of Conferences font
\usepackage{listings}
\usepackage{hyperref}
\begin{document}
\title{Variable stars in the northern Galactic plane from KISOGP}

\author{\firstname{Noriyuki} \lastname{Matsunaga}\inst{1}\fnsep\thanks{\href{mailto:matsunaga@astron.s.u-tokyo.ac.jp}{\tt matsunaga@astron.s.u-tokyo.ac.jp}} 
on behalf of KISOGP team
}

\institute{Department of Astronomy, The University of Tokyo, 7-3-1 Hongo, Bunkyo-ku, Tokyo 113-0033, Japan}

\abstract{%
We have conducted a large-scale survey of the northern plane using
Kiso Wide Field Camera attached to Schmidt telescope at Kiso observatory.
The KISOGP (KWFC Intensive Survey of the Galactic Plane) project have made
40--70 epoch observations in $I$ band of about 320~{sq.} degrees for
5 years starting in 2012. The limiting magnitude is around $16.5$ in $I$.
In the data analysis so far, we detected a couple of thousands of
variable stars including approximately 100 Cepheids and more than 700 Miras.
Roughly 90\% of them were not previously reported as variable stars,
indicating that there are still many relatively bright variables to be found
in the Galactic plane.
%% We'll present a preliminary map of the Cepheids and Miras together
%% with results of spectroscopic follow-up observations for these objects.
}
\maketitle

A large fraction of stars and gas is concentrated within
the Galactic disk. In fact, approximately half of the baryonic mass
within our Galaxy, ${\sim}4 \times 10^{10}~M_\odot$, is found in the disk
(\cite{Bland-Hawthorn-2016}). 
The stellar populations in a large space of the disk have, however,
remain elusive due to the interstellar extinction. Historic surveys
of variable stars, mostly done in the optical wavelengths,
were far from complete (see, e.g.\  figure~1 in \cite{Matsunaga-2012}).
{\it Gaia} and other ongoing surveys in the optical are
expected to find many new variables, while infrared surveys would be
more effective for objects obscured in the disk
(see our review \cite{Matsunaga-2017} in this proceedings).
We have conducted an $I$-band survey of variable stars in
the northern Galactic disk because it has been less explored
compared to the southern part for which
the OGLE\footnote{http://ogle.astrouw.edue.pl} and
the VVV\footnote{https://vvvsurvey.org} have been 
discovered a large number of new variables.

The instrument used for our survey is Kiso Wide Field Camera (KWFC) 
attached to 105-cm Schmidt telescope in Kiso observatory, Japan.
This camera has been designed for wide-field
observations by taking the advantage of a large focal-plane area of
the Schmidt telescope. Eight CCD chips with a total of 64~Mpixels
cover a field-of-view of $2.2^\circ \times 2.2^\circ$.
The distortion is negligible (at least smaller than 0.3
$0.3^{\prime\prime}$) across the entire 4.8~deg$^2$ field.
The survey is named KWFC Intensive Survey of the Galactic Plane (KISOGP)
and conducted for 5 years starting in 2012.
The main goal is to study the Galactic structure using pulsating stars
as tracers and we have been also conducting follow-up observations
of variable stars we find in both photometric and spectroscopic ways.

\begin{figure}
\centering
\includegraphics[width=13.5cm,clip]{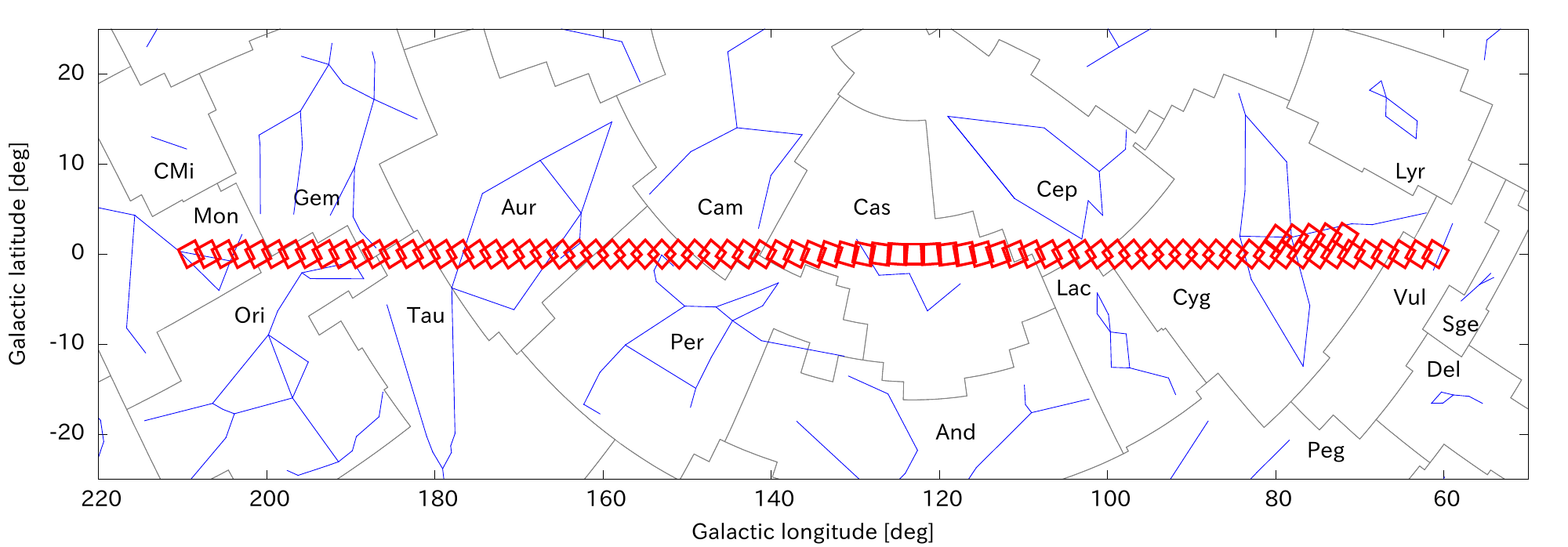}
\caption{KISOGP survey region.}
\label{fig:regions}       % Give a unique label
\end{figure}

Our survey region is composed of 80 KWFC fields-of-view
covering ${\sim 320}$~deg$^2$ between 60 and 210~degrees
in Galactic longitude (Figure~\ref{fig:regions}).
The photometric observations are only in $I$ band
and cover the magnitude range of 9.5 to 16.5~magnitude in $I$
with the signal-to-noise ratio better than 30.
For each epoch per each field, an exposure with 5~sec
and three exposures with 60~sec are combined.
The numbers of epochs for the 5-yr time monitoring ranges from
50 to more than 100 of which 40- or more-epoch datasets are
confirmed to be of useful quality.

In our entire survey field, we have detected more than 5 million objects,
of which about 50,000 candidates of variable stars are found,
and data analysis is still on progress.
So far we have searched for periodic variable stars with relatively
large amplitudes and identified ${\sim}100$~Cepheids and more than 700 Miras.
Roughly 90~\% of them were not reported as variables before.
Preliminary estimates of their distances using the period-luminosity relations
and 2MASS $JHK_{\rm s}$ magnitudes indicate that they are scattered across
a large space of the northern Galactic disk, from a couple of kpc
to more than 10~kpc from the Sun. Their distribution of the Galactocentric
distances covers ${\sim}8$ to 15~kpc. They will be important tracers of
the outer disk.
There are also RR Lyrs, a large number of eclipsing binaries ($>1000$), and
other types of variables. While we are trying to finalize the data analysis
for these variables, there are many other variables with smaller amplitudes
and/or irregular variations.

In order to take advantage of the newly found Cepheids and Miras,
we have been conducting several follow-up observing programs.
For Cepheids, we have used Subaru telescope to take high-resolution spectra
and several telescopes including
Las Cumbres Observatory's global network\footnote{https://lco.global} and
to collect multi-color photometric data.
For Miras, we have collected hundreds of low-resolution spectra
to classify them into oxygen-rich and carbon-rich Miras
using 188-cm telescope at Okayama Astrophysical
Observatory\footnote{http://www.oao.nao.ac.jp/en/}, NAOJ,
and 2-m Nayuta telescope at Nishi-Harima Astronomical
Observatory\footnote{http://www.nhao.jp/en/}.

\begin{acknowledgement} 
\noindent\vskip 0.2cm
\noindent {\em Acknowledgments}: We acknowledge a long-term support
for the KISOGP project from the staff in Kiso observatory. 
NM appreciates financial support from
the Japan Society for the Promotion of Science (JSPS) through
the Grant-in-Aid, No.~26287028.
\end{acknowledgement}

\end{document}